\documentclass[aps,pre,twocolumn,showpacs,preprintnumbers,amsmath,amssymb]{revtex4-1}

\begin{document}


\preprint{}

\title{Casimir force for geometrically confined ideal Bose gas in a harmonic-optical potential}


\author{Ekrem Aydiner}
\email[E-Mail: ]{ekrem.aydiner@istanbul.edu.tr}
\affiliation{Department of Physics, Istanbul University, TR-34134 Istanbul, Turkey}

%

\date{\today}

\begin{abstract}
In this study, we have derived close form of the Casimir force for
the non-interacting ideal Bose gas between two slabs in
harmonic-optical lattice potential by using Ketterle and van Druten
approximation. We find that Bose-Einstein condensation temperature
$T_{c}$ is a critical point for different physical behavior of the
Casimir force.  We have shown that Casimir force of confined Bose
gas in the presence of the harmonic-optical potential decays with
inversely proportional to $d^{5}$ when $T\leq T_{c}$. However, in
the case of $T>T_{c}$, it decays exponentially depends on separation
$d$ of the slabs. Additionally we have discussed temperature
dependence of Casimir force and importance of the harmonic-optical
lattice potential on quantum critical systems, quantum phase
transition and nano-devices.
\end{abstract}



\maketitle

\section{Introduction}
\label{intro}

The Casimir effect is an attractive or repulsive long-range
interaction between conducting boundaries \cite{Casimir1948}, which
originates from the quantum mechanical fluctuations of the
electromagnetic field in vacuum. The electromagnetic field is
confined in between plates with distance $d$, the fluctuations modes
of electromagnetic fields need to have a node at the plate surfaces,
so that only waves with wavelength $\lambda=nd/2$ and integer $n$
are permitted in the gap between plates. As a consequence the total
field outside the gap produces a pressure on the plates which is
higher than the one produced from inside, so the surfaces are pushed
together by force This effect has been experimentally measured in
1997 \cite{Lamoreaux1997} and 1998 \cite{Mohideen1998} which
confirms quantum field theory. It is shown that the force depends on
size, temperature, geometry, surface roughness and electronic
properties of the materials
\cite{Milton2001,Mostepanenko1997,Bordag2001}. Its importance for
practical applications is now becoming more widely appreciated in
quantum field theory, gravitation and cosmology, Bose-Einstein
condensation (BEC), atomic and molecular systems, mathematical
physics and nano-technology. A good review on new developments in
the Casimir effect can be found in Ref.\,\cite{Bordag2001}.

Now it is well-known that the fluctuations which can emerge from
different physical mechanisms can lead to Casimir-like
effect if it confined in the boundaries. For example, the thermal fluctuation in confined
quantum critical systems can leads to Casimir-like effect. This
emerges upon approaching a second-order phase transition point which
is characterized by the fact that due to the emerging collective
behavior of the particles depends on thermal fluctuation. The
effective force resulting from the confinement of the fluctuations
of the order parameter is called Casimir force. This phenomena
firstly was predicted by Fisher and de Genes \cite{Fisher1978}. They
showed that the fluctuation-induced force arise in critical systems
which are close to the critical point, where dynamics are governed
by the thermal fluctuations of the order parameter upon approach to
the thermal phase transition. This effect later is called as the
critical Casimir effect for quantum particle systems
\cite{Krech1994}. Firstly in Ref.\,\cite{Martin2006}, the Casimir
force between two slabs immersed in a perfect gas was calculated for
various boundary conditions. It was found that the Casimir force has
the standard asymptotic form with universal Casimir terms below the
BEC critical temperature $T_c$ and vanishes exponentially above
$T_c$. After these seminal works, a great deal of effort has been
devoted to calculation of the Casimir force of the free and trapped
Bose gas for different geometries and different boundary conditions
\cite{Gambassi2006,Biswas2007a,Biswas2007b,Gambassi2009,
Napiorkowski2011,Napiorkowski2012,Napiorkowski2013,
Lin2012,Biswas2010,Dantchev2003,Roberts2005,Edery2006,Yu2009,Hucht2007,
Vasilyev2007,Maciolek2007,Hasenbusch2009,Hasenbusch2010,Garcia1999,
Ganshin2006,Aydiner2013}.

Another mechanism of Casimir-like effect arises from quantum size
effect. It is expected that Casimir-like effects may occur in
geometrically confined quantum gas because of the wave character of
the gas atoms at nano scale. Indeed it is shown that in the presence
of boundaries inside quantum particles can cause Casimir-like effect
if the bounded space is smaller than the maximum thermal de Broglie
wavelength of the particles
\cite{Molina1996,Pathira1998,Dai2003,Dai2004,
Sisman2004,Pang2006,Firat2013,Nie2008a,Nie2008b,Nie2009a,Nie2009b,Nie2010}.
Therefore, nowadays it has been attracted to Casimir force at nano
scale because of the potential applications in nano-technology area
since Casimir phenomena is a typically quantum size effect which
appears at nano scale. In this area, researchers aim to produce many
nano-device structures such as gas turbines, pumps, mixers, heat
exchangers, valves, etc
\cite{Nie2008a,Nie2008b,Nie2009a,Nie2009b,Nie2010}. Therefore
understanding the Casimir effect on the thermodynamic behavior of
quantum gases confined within boundaries at nano scale in
nano-devices is very significant problem. Because new nano-devices
and technologies can be developed based on the effects appearing at
this scale.

The subject of Casimir effect on thermodynamic behaviors of gases at
nano scale and in quantum critical systems is a new research area
and may have many potential applications in nano technology and
other area in physics. In this regard, discussing on the Casimir
effect in the previous theoretical studies in
Refs.\,\cite{Fisher1978,Krech1994,Martin2006,Gambassi2006,Biswas2007a,Biswas2007b,Gambassi2009,
Napiorkowski2011,Napiorkowski2012,Napiorkowski2013,
Lin2012,Biswas2010,Dantchev2003,Roberts2005,Edery2006,Yu2009,Hucht2007,
Vasilyev2007,Maciolek2007,Hasenbusch2009,Hasenbusch2010,Garcia1999,
Ganshin2006,Aydiner2013,Molina1996,Pathira1998,Dai2003,Dai2004,Sisman2004,Pang2006,Firat2013,Nie2008a,Nie2008b,Nie2009a,Nie2009b,Nie2010}
have potential to enlighten the thermodynamic behavior of confined
quantum gases. So far Casimir force has been calculated for either
free or trapped Bose gas with harmonic potential
\cite{Martin2006,Gambassi2006,Biswas2007a,Biswas2007b,Gambassi2009,
Napiorkowski2011,Napiorkowski2012,Napiorkowski2013,
Lin2012,Biswas2010,Dantchev2003,Roberts2005,Edery2006,Yu2009,Hucht2007,
Vasilyev2007,Maciolek2007,Hasenbusch2009,Hasenbusch2010,Garcia1999,
Ganshin2006,Aydiner2013}. As it is known that the harmonic-optical
potential which consists of combining harmonic and optical lattice
potentials plays very important role in BEC systems as well as in
other area of physics. The effects of harmonic-optical lattice
potential on the BEC and its vortices and quantum phase transition
have been considered in Refs.\,
\cite{Greiner2002,Bargi2006,Cozzini2006,Fetter2001,Ghosh2004,
Kim2005,Danaila2005,Fetter2005,Kling2007,Blakie2007,
Dupuis2009,Gerbier2008,Hassan2010}. But to our knowledge Casimir
force of Bose gas trapped with harmonic-optical potential has never
been studied. In this study we investigate Casimir force of the Bose
gas trapped with harmonic-optical potential between two infinite
parallel slabs in the $x-y$ direction that are separated by a
distance $d$ in the $z$-direction. We obtain explicit form of the
Casimir force of the ideal Bose gas trapped in harmonic-optical
potential and we show that Casimir force decays with inversely
proportional to $d^{5}$ when $T\leq T_{c}$ while it decays
exponentially for $T > T_{c}$.

\section{Model and Analytical Results}

\subsection{Theoretical framework}
In the original paper \cite{Casimir1948} Casimir force of vacuum
fluctuation of electromagnetic field at zero temperature ($T=0$) was
defined as
\begin{eqnarray}
F_{C}=-\frac{\partial}{\partial d}\left[ E\left(  d\right) -E\left(
\infty\right)  \right]
\end{eqnarray}
where $E\left(  d\right)$ is the ground-state energy (i.e. the
vacuum energy) of the electromagnetic field in between the two
conducting plates separated at a distance $d$.  $E\left(
\infty\right)$ is the energy at infinite. We consider the Casimir
effect for the thermodynamical system of Bose gas between two
infinite slabs. The geometry of the system on which some external
boundary condition can be imposed is responsible for the Casimir
effect. In recent studies it has been shown that the Casimir force
for critical particle systems can be obtained from Casimir potential
at finite temperature
\cite{Martin2006,Gambassi2006,Biswas2007a,Biswas2007b,
Gambassi2009,Roberts2005,Edery2006,Yu2009,Napiorkowski2012,
Napiorkowski2011,Lin2012,Biswas2010,Napiorkowski2013,Dantchev2003}
as
\begin{eqnarray}
F_{C}\left(  T,\mu,d\right)=-\frac{\partial}{\partial d}\left[
\varphi_{C}\left(  T,\mu,d\right)  - \varphi_{C}\left(  T,\mu,\infty \right)\right]
\end{eqnarray}
where $\varphi_{C}\left(  T,\mu,d\right)$ and $\varphi_{C}\left(
T,\mu,\infty \right)$ are the Casimir potential corresponds to
fluctuations of thermally excited states of Bose gas at distance $d$
and at infinite, respectively
\cite{Martin2006,Gambassi2006,Krech1994,Biswas2007a,Biswas2007b,Gambassi2009}.
The Casimir potentials in Eq.\,(2) can be obtained from grand
canonical potential of the particle system
\begin{eqnarray}
\varphi\left(  T,\mu,d\right) =-\beta^{-1}\sum_{n}^{\infty}\ln\left\{
1-e^{-\beta\left( \varepsilon_{n} -\mu\right)
}\right\}
\end{eqnarray}
where $\varepsilon_{n}$ is the energy of Bose gas, which corresponds to
energy eigenvalue of the particles. Additionally $\beta=\left(kT\right)^{-1}$, $k$ is
Boltzmann's constant, $T$ is temperature, and $\mu$ is the chemical
potential of the system. For simplicity we consider single particle
energy instead of $N$ particle Bose system.

\subsection{Single particle energy}

In present model, we assume that ideal Bose gas is trapped in two
dimensional  harmonic-optical potential between two slabs separated
at a distance $d$. If quantum particles move in three dimensional
space in the absence external potential, single particle energy at
finite temperature can be given by $\varepsilon\left(
p_{x},p_{y},p_{z}\right) =p_{x}^{2}/2m + p_{y}^{2}/2m +
p_{z}^{2}/2m$ where $m$ is the mass of the particle, and $p_{x}$,
$p_{y}$ and $p_{z}$ are the momentum along respectively $x$, $y$ and
$z$ axis. However in our model particles are trapped in between
quasi-two-dimensional slabs with two dimensional potential. Hence
the energy eigenvalues of particles at $x$ and $y$-directions depend
on the harmonic-optical potentials. For a simplicity, it can be
assumed for present model that particle energy $\varepsilon$ depends
on quantum number $n_{x,y}$ in $x$ and $y$-directions and in the
$z$-direction, energy of the particle is quantized as
$\varepsilon_{n_{z}}=(a\pi^{2}\hbar^{2}n_{z}^{2})/(2md^{2})$ where
$n_{z}=0,1,2,3,...$\, and $a$ indicates type of the boundary
condition of the system. To define boundary conditions, it can be
set as $a=1$ and $n_{z}=1,2,3,...$ for Dirichlet, $a=1$ and $
n_{z}=0,1,2,...$ for Neumann, and $a=2$ and $n_{z}=0,\pm1,\pm2,...$
for periodic. Therefore, the single particle energy for this model
is arranged in the form
\begin{eqnarray}
\varepsilon_{n_{x,y,z}}=\varepsilon_{n_{x,y}}+ \frac{a\hbar^{2}\pi^{2}}{2md^{2}}n_{z}^{2} \ .
\end{eqnarray}
where $\varepsilon_{n_{x,y}}$ corresponds to two dimensional quantized energy of the particle in the $x-y$ directions, $\varepsilon_{n_{x,y,z}}$ corresponds to energy of the particles for all degrees of freedom. In order to obtain grand canonical potential of an trapped ideal Bose gas we have to compute single particle energy in Eq.\,(4) for the Bose gas.

The first term of rhs in Eq.\,(4) must be determined to obtain exact
expression for single particle energy $\varepsilon_{n_{x,y,z}}$. The
particle energy $\varepsilon_{n_{x,y}}$ in $x-y$ plane can be
determined depending on harmonic-optical potential
\begin{eqnarray}
V\left(  x,y\right)  =V_{har}\left(  x,y\right)
 +V_{lat}\left(
x,y\right)
\end{eqnarray}
where harmonic potential and optical lattice potential are respectively given
\begin{eqnarray}
V_{har}\left(  x,y\right)  =\frac{1}{2}m\left(  \omega_{x}^{2}x^{2}%
+\omega_{y}^{2}y^{2} \right)
\end{eqnarray}
\begin{eqnarray}
V_{lat}\left(  x,y\right)  =V_{0}\left(  \sin^{2}kx+\sin^{2}ky
\right)
\end{eqnarray}
With $m$ is the mass of the particle, $\left\{
\omega_{x},\omega_{y}\right\}$ are the frequencies of the harmonic
potential along the coordinate directions $x$ and $y$. $V_{0}$ is
the lattice potential depth and $k=2\pi/\lambda^{\prime}$ is the
wave vector of the laser light, $\lambda^{\prime}$ is the laser
wavelength. In terms of $k$ we can write the recoil energy
$\varepsilon_{R}=\hbar^{2}k^{2}/2m$ $\left\{
\equiv\hbar\omega_{R}\right\}$ as an energy scale for specifying the
lattice depth. It is defined as the recoil energy that one atom
requires when it absorbs one lattice photon \cite{Hassan2010}.

For the potential (5) it is impossible to find an exact
analytical expression for the energy levels $\varepsilon_{n_{x,y}}$.
An approximate
expression can be readily obtained by extending the standard
harmonic approximation for the optical lattice potential
\cite{Blakie2007,Hassan2010}. Taylor expansion of Eq.\,(7) about the lattice site minimum at
$x=y=0$ leads to
\begin{eqnarray}
V_{lat}\left(  x,y\right)  =\frac{V_{0}}{2}\left\{  \left(
1-\cos2kx\right) +\left(  1-\cos2ky\right) \right\}
\end{eqnarray}
which is approximately equal to
\begin{eqnarray}
V_{lat}\left(  x,y\right)  =V_{0}k^{2}\left(  x^{2}+y^{2}\right)
-\frac{V_{0}}{2}k^{4}\left(  x^{4}+y^{4} \right) \ .
\end{eqnarray}
As it can be seen from Eq.\,(9) that the first term can be organized
in the form of the harmonic oscillator potential
$\frac{1}{2}m\omega_{lat}^{2}\left(  x^{2}+y^{2}\right)$ which
yields an effective harmonic oscillator frequency
\begin{eqnarray}
\omega_{lat}=\sqrt{\frac{2V_{0}k^{2}}{m}}=\frac{2\sqrt{E_{R}V_{0}}}{\hbar}
\end{eqnarray}
with equivalent harmonic oscillator length
$l=\sqrt{\hbar/m\omega_{lat}}$. In this case the localized states in
the optical lattice can be approximated by a harmonic oscillator
states, $\hbar\omega_{lat}$. Under this consideration the effective
confining potential can be given by
\begin{widetext}
\begin{eqnarray}
V_{eff}\left(  x,y \right)  =\frac{1}{2}m\left[  \left(  \omega_{x}^{2}%
+\omega_{lat}^{2}\right)  x^{2}+\left(  \omega_{y}^{2}+\omega_{lat}%
^{2}\right)  y^{2} \right] -\frac{V_{0}}{2}k^{4}\left( x^{4}+y^{4}
\right) \ .
\end{eqnarray}
\end{widetext}
This potential is characterized by a single particle energy levels
with Dirichlet boundary condition given by
\begin{eqnarray}
\varepsilon_{n_{x,y}}=n_{x}\hbar\omega_{x}^{\prime}+n_{y}\hbar\omega_{y}^{\prime
}+\hbar\overline{\omega}_{1}-\Delta\varepsilon_{n}
\end{eqnarray}
where $\omega_{i}^{\prime}=\sqrt{\omega_{i}^{2}+\omega_{lat}^{2}}$
with $i$ stands for $x$ and $y$, on the other hand,
$n_{x},n_{y}=0,1,2,...$ and $\overline{\omega}_{1}=\frac{1}{2}\left(
\omega_{x}^{\prime}+\omega _{y}^{\prime}\right)$ is the mean of
combined frequencies. In deriving Eq.\,(12), the second term in
Eq.\,(11), quartic term is treated perturbativly by using the normal
ladder operators \cite{Blakie2007}. This term provides a shift in
the oscillator state energies. Up to the first order term this shift
is given by
\begin{eqnarray}
\Delta\varepsilon_{n}=\frac{\hbar\omega_{R}}{2}\left[  \left(  n_{x}%
^{2}+n_{y}^{2}\right)  +\left(  n_{x}+n_{y}\right)
+1\right]
\end{eqnarray}
where $\omega_{R}=\hbar k^{2}/2m$ is the recoil frequency. Note
that for a fixed value of $\varepsilon_{R}=\hbar\omega_{R}$ the
quantity $\Delta\varepsilon_{n}$ grows rapidly with $n^{\prime}$s. This is
not surprising, since the higher excited states of the harmonic
oscillator extend to larger and larger values of $x$ and $y$. In
such case the perturbation term can have important effect for higher
excited state. Also from Eq.\,(13) we see that the higher the value
of $n_x$ and $n_y$ are smaller than the values of $\varepsilon_{R}$ for
which reliable values may be obtained from the experimental set-up
conditions, $\varepsilon_{R}=\hbar^{2}k^{2}/2m$, where $k$ is
determined from the laser beam wavelength. Thus in order to keep the
value of $\varepsilon_{R}$ in its experimental range one has to
neglect the term which contains $\left(  n_{x}^{2}+n_{y}^{2}\right)$
in Eq.\,(13). Under this approximation the single particle energy
level depends on  $n_x$ and $n_y$ is given
\begin{eqnarray}
\varepsilon_{n_{x,y}}=n_{x}\hbar\left(  \omega_{x}^{\prime}-\frac{\omega_{R}}%
{2}\right)  +n_{y}\hbar\left(  \omega_{y}^{\prime}-\frac{\omega_{R}}%
{2}\right) +\varepsilon_{0}
\end{eqnarray}
where $\varepsilon_{0} = \hbar\overline{\omega}$ is the
single particle ground state energy with
$\overline{\omega}=\overline{\omega}_{1}-\frac{\omega_{R}}{2}$ is
the meaning of the combined frequencies. Eq.\,(14) means that particle oscillates in quasi-two dimensional space. On the other hand, there is an energy contribution to particle in the $z$-direction. For Dirichlet boundary condition this contribution can be given as  $\frac{\hbar^{2}\pi^{2}}{2md^{2}}n_{z}^{2}$.  Hence
the single particle energy in all directions can be defined by
\begin{widetext}
\begin{eqnarray}
\varepsilon_{n_{x,y,z}}=n_{x}\hbar\left(  \omega_{x}^{\prime}-\frac{\omega_{R}}%
{2}\right)  +n_{y}\hbar\left(  \omega_{y}^{\prime}-\frac{\omega_{R}}%
{2}\right) +\frac{\hbar^{2}\pi^{2}}{2md^{2}}n_{z}^{2}+\varepsilon_{0}
\end{eqnarray}
Now with help Eq.\,(15) grand canonical potential for two
dimensional trapped Bose gas in harmonic-optical potential in
between two slabs can be obtained.

\subsection{Thermodynamical Potential}
Grand canonical potential (3) can be written as
\begin{eqnarray}
\varphi\left(  T,\mu,d\right)  = -\varphi_{0}  - kT\sum
_{n=1}^{\infty}\ln\left(  1-ze^{-\beta\varepsilon_{n_{x,y,z}}}\right)
\end{eqnarray}
where first term $\varphi_{0}$ is ground state potential and second term corresponds to the potential of excited particles. We have neglected the ground-state term which does not give any contribution to Casimir potential at finite temperature since the Casimir effect was caused by thermal or near-critical fluctuations of the order parameter upon approaching the condensation transition.
The second term in Eq.\,(16) for finite particle can be presented as
\begin{eqnarray}
\varphi\left(  T,\mu,d\right)  = -kT\sum_{n=1}^{\infty}\sum_{j=1}^{\infty
}\frac{z^{j}}{j}e^{-j\beta\varepsilon_{n_{x,y,z}}}
\end{eqnarray}
If Eq.\,(15) is put into Eq.\,(17), grand canonical potential for two dimensional trapped Bose gas can be written in generalized form
\begin{equation}
\varphi\left(  T,\mu,d\right)=
-kT\sum_{n_{x}}^{\infty}\sum_{n_{y}}^{\infty}\sum_{n_{z}=1}^{\infty}\sum_{j=1}^{\infty}
\frac {z^{j}}{j}\exp\left\{
-j \beta \left(  n_{x}\hbar\left(  \omega_{x}^{\prime}-\frac{\omega_{R}}%
{2}\right)  +n_{y}\hbar\left(  \omega_{y}^{\prime}-\frac{\omega_{R}}%
{2}\right)  +\frac{\hbar^{2}\pi^{2}}{2md^{2}}n_{z}^{2}\right)
\right\}
\end{equation}
where $z=e^{\beta\mu}$. Here we have neglected the energy
$\varepsilon_{0}$ in Eq.\,(15) since ground state does not
contribute to Casimir potential. Eq.\,(18) can be simplified in the
form
\begin{eqnarray}
\varphi\left(  T,\mu,d\right)= -kT\sum_{n_{z}=1}^{\infty}\sum_{j=1}^{\infty}\frac{\frac {z^{j}}{j}\exp\left\{  -j \beta \left(
\frac{\hbar^{2}\pi^{2}}{2md^{2}}n_{z}^{2}\right)  \right\}  }{\left(
1-\exp\left\{ j \beta \hbar\left(  \omega_{x}^{\prime}-\frac{\omega_{R}%
}{2}\right)  \right\}  \right)  \left(  1-\exp\left\{ j \beta
\hbar\left(  \omega_{y}^{\prime}-\frac{\omega_{R}}{2}\right)  \right\}
\right)  }
\end{eqnarray}
To evaluate the sum over $j$, we need an approximation due to Ketterle and van Druten \cite{Ketterle1996} which yields near exact results. This approximation uses the thermodynamic limit $\hbar\omega\ll kT$ to write
$\left(  1-\exp\left\{ j \beta \hbar\left(  \omega_{i}^{\prime}
-\frac{\omega_{R}}{2}\right)  \right\}  \right)  =j\beta \hbar\left(
\omega_{i}^{\prime}-\frac{\omega_{R}}{2}\right)$ where $i=x,y$. The sum over $j$ has the factor $1/j^{2}$ and is convergent. Thus, Eq.\,(19) is written as
\begin{eqnarray}
\varphi\left(  T,\mu,d\right)  = -kT\frac{\left(  \frac{kT}{\hbar}\right)
^{2}}{\left(  \omega_{x}^{\prime}-\frac{\omega_{R}}{2}\right)  \left(
\omega_{y}^{\prime}-\frac{\omega_{R}}{2}\right)  }
\sum_{n=1}^{\infty}\sum_{j=1}^{\infty}%
\frac{z^{j}}{j^{3}}\exp\left\{  -j \beta \left(
\frac{\hbar^{2}\pi^{2}}{2md^{2}}n^{2}\right)  \right\}  \ .
\end{eqnarray}
where we set $n_{z}=n$ for simplicity. This analytical expression
corresponds to the grand canonical potential of trapped Bose gas in
quasi-two-dimensional harmonic-optical potential in between two
slabs due to Ketterle and van Druten approximation.
\end{widetext}
\subsection{Casimir potential}
Casimir potential can be obtained from Eq.\,(20). However, Eq.\,(20)
includes contributions of the bulk, surface and fluctuation of
excited states of the Bose gas between two slabs separated by $d$.
Bulk and surface potentials do not contribute to the Casimir force,
because the force due to the bulk term is counterbalanced by the
same contribution acting from outside the slabs when they are
immersed in the critical medium
\cite{Lin2012,Martin2006,Gambassi2009}, on the other hand, the
surface potential caused from two slab-boson gas interfaces, does
not change with changing $d$. However thermal fluctuations of
excited states of Bose gas leads to Casimir potential. Therefore, to
obtain Casimir potential, these contributions must be separated into
components. By using the Jacobi identity \cite{Hunter2000}
\begin{eqnarray}
\sum_{n=1}^{\infty}e^{-\pi n^{2}b}=\left(  \frac{1}{2\sqrt{b}}-\frac{1}%
{2}\right)  +\frac{1}{\sqrt{b}}\sum_{n=1}^{\infty}e^{-\pi n^{2}/b}
\end{eqnarray}
where $b=j\pi\left(\lambda/d\right)^{2}/2$ with $b>0$ and $\lambda=\hbar\sqrt{\beta/m}$ is thermal de Broglie wavelength of the particles, these contributions can be separated as
\begin{widetext}
\begin{eqnarray}
\varphi\left(  T,\mu,d\right) = - \frac{\left(  kT\right)  ^{3}}{\hbar^{2}\left(  \omega_{x}^{\prime}%
-\frac{\omega_{R}}{2}\right)  \left(  \omega_{y}^{\prime}-\frac{\omega_{R}}%
{2}\right)} \sum_{j=1}^{\infty}%
\frac{z^{l}}{j^{3}}\left[ \left(  \frac{1}{2\sqrt{b}}-\frac{1}%
{2}\right)  +\frac{1}{\sqrt{b}}\sum_{n=1}^{\infty}e^{-\pi n^{2}/b}\right] \ .
\end{eqnarray}

In Eq.\,(22) the first term corresponds to bulk potential
$\varphi_{bulk}\left(T,\mu,d\right)$ and second term is the surface
potential $\varphi_{surf}\left(  T,\mu,d\right)$.  However, the last
term in Eq.\,(22) corresponds to the Casimir potential for combined
potential
\begin{eqnarray}
\varphi_{C}\left(  T,\mu,d\right)= - \frac{\left(  kT\right)^{3}}{\hbar^{2}\left(  \omega_{x}^{\prime}%
-\frac{\omega_{R}}{2}\right)  \left(  \omega_{y}^{\prime}-\frac{\omega_{R}}%
{2}\right)} \sqrt{\frac{2}{\pi}}\frac{d}{\lambda}%
\sum_{j=1}^{\infty} \sum_{n=1}^{\infty}\frac{e^{j\beta\mu}}{j^{7/2}%
}e^{-2\left( nd/\lambda\right)^{2}/j} \ .
\end{eqnarray}
where sum over $j$ and $n$. The summation in right hand side of
Eq.\,(23) can be converted to integral in the limit $d/\xi\ll1$,
\begin{eqnarray}
\sum_{j=1}^{\infty}\sum_{n=1}^{\infty}\frac{e^{j\beta\mu}}{j^{7/2}%
}e^{-2\left(  nd/\lambda\right)  ^{2}/j}
=2\left(  \frac{\lambda}{d}\right)
^{5}\int_{0}^{\infty}x^{-6}e^{-px^{2}-q/x^{2}}dx \ .
\end{eqnarray}
where $p=u^{2}/2=\left(  -2\beta\mu\right)  ^{1/2}d/\lambda\sim d/\xi$, $\xi$ is the correlation length, $q=2n^{2}$ and $x^{2}=\left(  \frac{\lambda}{d}\right)^{2}r$. Thus, Casimir potential is
\begin{eqnarray}
\varphi_{C}\left(  T,\mu,d\right)  = - \frac{2\left(  kT\right)  ^{3}}{\hbar^{2}\left(  \omega_{x}^{\prime}%
-\frac{\omega_{R}}{2}\right)  \left(  \omega_{y}^{\prime}-\frac{\omega_{R}}%
{2}\right)}\sqrt{\frac{2}{\pi}}\left(  \frac{\lambda
}{d}\right)  ^{4}\int_{0}^{\infty}x^{-6}e^{-px^{2}-q/x^{2}}dx
\end{eqnarray}
After some mathematical algebra, Casimir potential can be obtained as follows
\begin{eqnarray}
\varphi_{C}\left(  T,\mu,d\right) =-\frac{3\left(  kT\right)  ^{3}}{64\hbar
^{2}\pi^{2}\left(  \omega_{x}^{\prime}-\frac{\omega_{R}}{2}\right)  \left(
\omega_{y}^{\prime}-\frac{\omega_{R}}{2}\right)  }\left(  \frac{\lambda}%
{d}\right)  ^{4}\sum_{n=1}^{\infty}\left(  \frac{1+2un+4u^{2}n^{2}/3}{n^{5}%
}\right)  e^{-2un} \ .
\end{eqnarray}
If $\lambda=\hbar\sqrt{\beta/m}$ is inserted into Eq.\,(26), Casimir potential is given in a simple form
\begin{eqnarray}
\varphi_{C}\left(  T,\mu,d\right)  =-\frac{3\hbar^{2}kT}{64\pi^{2}m^{2}\left(  \omega_{x}^{\prime
}-\frac{\omega_{R}}{2}\right)  \left(  \omega_{y}^{\prime}-\frac{\omega_{R}%
}{2}\right)  }\frac{1}{d^{4}}\sum_{n=1}^{\infty}\left(  \frac{1+2un+4u^{2}%
n^{2}/3}{n^{5}}\right)  e^{-2un}
\end{eqnarray}
which originates from fluctuations of thermally excited states of Bose gas in between slabs.
\end{widetext}
\subsection{Casimir Force}
\begin{widetext}
Now we can compute Casimir force for tapped ideal Bose gas in
harmonic-optical potential between two slabs by using of Eqs.\,(2)
and (27). Casimir potential in Eq.\,(27) goes to zero i.e.,
$\varphi_{C}\left(T,\mu,\infty\right)\rightarrow0$ for $d\rightarrow
\infty$. Therefore Casimir force in Eq.\,(2) reduces to simple form
\begin{eqnarray}
F_{C}\left(T,\mu,d\right)=-\frac{\partial}{\partial d}
\varphi_{C}\left(  T,\mu,d\right)
\end{eqnarray}
which implies that Casimir force of Bose gas changes due to the changing
of distance between slabs. From Casimir potential (27), an general analytical form for the force is given simply by
\begin{eqnarray}
F_{C}\left(T,\mu,d\right)= \frac{3kT\hbar^{2} }{16\pi^{2}m^{2}\left(  \omega_{x}^{\prime}-\frac
{\omega_{R}}{2}\right)  \left(  \omega_{y}^{\prime}-\frac{\omega_{R}}%
{2}\right)  }\frac{1}{d^{5}} \sum_{n=1}^{\infty}\left(  \frac{1+2un+4u^{2}%
n^{2}/3}{n^{5}}\right)  e^{-2un}
\end{eqnarray}
\end{widetext}
However this expression does draw a physical picture of the Casimir
force for the Bose gas since it depends on chemical potential $\mu$
of the gas. It is well known that $\mu$ has a curious behavior
depends on temperature. In the Bose gas, thermal fluctuations cause
second order phase transition which is called BEC occur at critical
temperature $T_{c}$. Chemical potential $\mu$ equal to zero at the
$T=T_{c}$ and $\mu< 0$ when $T> T_{c}$. It is known that correlation
length $\xi$ of the thermal fluctuations diverge to infinite at the
second order phase transition temperature $T=T_{c}$ where most of
the particles begin to collapse to ground state and occur
condensation. However contribution to Casimir potential does not
come from ground state, it is meaning that BEC does not cause
Casimir effect. The origin of the Casimir effect in the critical
particle systems is excited states. The correlation length $\xi$ of
thermal fluctuation of the excited states of the quantum gas is
finite when $T\gtrapprox T_{c}$. If the geometry is restricted to be
that of two parallel plates separated by distance $d$ in the
$z$-direction to be $k=\pi n/d$ (for the Dirichlet boundary
condition), correlated fluctuations become dominate and the free
energy of the Bose gas is altered in the mediate vicinity of the
critical point $d/\xi<1$ leading to the Casimir force. Therefore, it
is suggested that Casimir effect for critical particle systems is
caused by thermal or near-critical fluctuations of the order
parameter upon approaching the condensation transition. In this
picture, we assume that in the near $T\gtrapprox T_{c}$, the
chemical potential to be $\mu=0$. In this case, when $T\simeq
T_{c}$, $z=1$ behaves as the bulk critical point for Casimir force.
Hence for $\mu=0$, the sum in Eq.\,(29) takes $\zeta\left(  5\right)
=\sum_{n=1}^{\infty}\frac{1}{n^{5}}$. Therefore, the Casimir force
in the vicinity the BEC transition temperature $T_{c}$ is obtained
in terms of Rieamann's Zeta function as follow
\begin{eqnarray}
F_{C}\left(T,\mu,d\right)=-\frac{3kT\hbar^{2}\zeta\left(5\right) }{16\pi^{2}m^{2}\left(  \omega_{x}^{\prime}-\frac
{\omega_{R}}{2}\right)  \left(  \omega_{y}^{\prime}-\frac{\omega_{R}}%
{2}\right)  }\frac{1}{d^{5}}
\end{eqnarray}
where $\omega_{x}^{\prime}=\sqrt{\omega_{x}^{2}+\omega_{lat}^{2}}$
and $\omega_{y}^{\prime}=\sqrt{\omega_{y}^{2}+\omega_{lat}^{2}}$.
For isotropic case, the Casimir force in combined potential reduce to
\begin{eqnarray}
F_{C}\left(T,\mu,d\right)=-\frac{3kT\hbar^{2}\zeta\left(5\right) }{16\pi^{2}m^{2}\left(  \omega^{\prime}-\frac
{\omega_{R}}{2}\right)^2 }\frac{1}{d^{5}}
\end{eqnarray}
where $\omega^{\prime}=\sqrt{\omega^{2}+\omega_{lat}^{2}}$. It can
be seen that Casimir force in BEC phase or $T\simeq T_{c}$ decays
inversely proportional to $d^{5}$ for harmonic-optical potential as
well as in the case of harmonic potential. However, as it can be
seen from Eqs.\,(30) and also (31) Casimir force of Bose gas clearly
depends on frequencies of harmonic and optical lattice potentials.

On the other hand, in the case of $T>T_{c}$, the chemical potential
is $\mu\neq 0$ and the double sums in Eq.\,(23) for $d/\xi\gg1$
satisfy the condition
\begin{eqnarray}
\sum_{j=1}^{\infty}\frac{e^{\beta\mu j}}{j^{7/2}}\sum_{n=1}^{\infty}e^{-2\left(  nd/\lambda\right)  ^{2}/j}<\frac{\zeta\left(
7/2\right)  }{e^{\sqrt{-8\mu}d/\lambda}-1} \ .
\end{eqnarray}
Therefore in the limit of large $d$, Casimir potential for
harmonic-optical potential can be given
\begin{eqnarray}
\varphi_{C}\left(T,\mu,d\right) \simeq   e^{-\sqrt{-8\mu
}d/\lambda} \ .
\end{eqnarray}
As it can be expected that for $d/\xi\gg1$ the Casimir potential has
exponential form which indicates that Casimir force also decays
exponentially with $d$. Additionally, at the high-temperature limit;
the correlated fluctuations between the two slabs are no longer
long-ranged therefore the confining boundaries are subject to a
vanishing Casimir force \cite{Gambassi2006,Lin2012}.

\section{Conclusion}

In this study, we have derived close form of the Casimir force for
the ideal Bose gas between two slabs in harmonic-optical potential
using by Ketterle and van Druten approximation. We have shown that
Casimir force of confined Bose gas in the presence of the
harmonic-optical potential decays with inversely proportional to
$d^{5}$ when $T\leq T_{c}$. However, in the case of $T>T_{c}$, it
decays exponentially depending on separation $d$ of the slabs.
Analytical results indicate that BEC critical temperature $T_{c}$
plays important role on the defining of the Casimir force in
confined critical particle systems. As discussed above, when
fluctuations are long ranges (comparable to smallest dimension of
the system) near the BEC transition in Bose gas there will be a
Casimir force between the confining surfaces. Based  on above
discussing we can suggest that similar behavior may happen for any
confined system near a critical point if the order parameter has
critical fluctuations on a scale $\xi$ which becomes large,
diverging in the bulk/thermodynamic limit. Finally, we state that
the role of the Casimir effect in confined systems such as BEC,
quantum phase transition at ultra cold Bose systems and nano
structures is still open problems. Harmonic-optical lattice
potential has a significant role on these systems. We hope that the
present analytical results may contribute to theoretical and
experimental studies in these area.


\vspace{0.2cm} \textit{Acknowledgments} -- The author would like to
thank to Professor Orhan Gemikonakli and to Department of Computer
and Communication Engineering School of Technology and Sciences
Middlesex University (London, UK) for hospitality during this work
in progress. This work has been supported by Istanbul University
under grant number No.28432 and 45662.


\begin{thebibliography}{99}

\bibitem{Casimir1948} H. B. G. Casimir, Proc. K. Ned. Akad. Wet. \textbf{51}, 793 (1948).

\bibitem{Lamoreaux1997} S. K. Lamoreaux, Phys. Rev. Lett. \textbf{81}, 5475 (1997).

\bibitem{Mohideen1998} U. Mohideen and A. Roy, Phys. Rev. Lett. \textbf{81}, 4549 (1998).

\bibitem{Milton2001} K. A. Milton, World Scientific, The Casimir Effect: Physical Manifestation of Zero-Point Energy, World Scientific, New Jersey (2001).

\bibitem{Mostepanenko1997} V. M. Mostepanenko, N. N. Trunov, The Casimir Effect and its Applications, Clarendon Press, Oxford, 1997.

\bibitem{Bordag2001} M. Bordag, U. Mohideen and V. M. Mostepanenko, Phys. Rep. \textbf{353}, 1 (2001).

\bibitem{Fisher1978} M. E. Fisher and P.G. de Gennes, C. R. Seances, Acad. Sci. Paris Ser. B \textbf{287}, 207 (1978).

\bibitem{Krech1994} M. Krech, The Casimir Effect in Critical Systems (World Scientific, Singapore, 1994).

\bibitem{Martin2006} P. A. Martin and V. A. Zagrebnov, Europhys. Lett. \textbf{73}, 15 (2006).

\bibitem{Gambassi2006} A. Gambassi and S. Dietrich, Europhys. Lett. \textbf{74}, 754 (2006).

\bibitem{Biswas2007a} S. Biswas, Euro. Phys. J. D. \textbf{42}, 109 (2007).

\bibitem{Biswas2007b} S. Biswas, J. Phys. A: Math. Theor. \textbf{40}, 9969 (2007).

\bibitem{Gambassi2009} A. Gambassi, J. Phys. A: Conf. Ser., \textbf{161}, 012037 (2009).

\bibitem{Napiorkowski2011} M. Napiorkowski and J. Piasecki, Phys. Rev. E \textbf{84}, 061105 (2011).

\bibitem{Napiorkowski2012} M. Napiorkowski and J. Piasecki, J. Stat. Phys. \textbf{147}, 1145 (2012).

\bibitem{Napiorkowski2013} M. Napiorkowski P Jakubczyk and K Nowak, J. Stat. Mech. P06015 (2013).

\bibitem{Lin2012} T. Lin, G. Su, Q. A. Wang and J. Chen, Europhys. Lett. \textbf{98}, 40010 (2012).

\bibitem{Biswas2010} S. Biswas, J. K. Bhattacharjee, D. Majumder1, K. Saha
and N. Chakravarty, J. Phys. B: At. Mol. Opt. Phys. \textbf{43} 085305 (2010).

\bibitem{Dantchev2003} D. Dantchev, M. Krech, S. Dietrich, Phys. Rev. E \textbf{67}, 066120 (2003).

\bibitem{Roberts2005} D. C. Roberts and Y. Pomeau, Phys. Rev. Lett. \textbf{95}, 145303 (2005).

\bibitem{Edery2006} A. Edery, J. Stat. Mech., P06007 (2006).

\bibitem{Yu2009} X. Yu, R. Qi, Z. B. Li and W. M. Liu, Europhys. Lett. \textbf{85}, 10005 (2009).

\bibitem{Hucht2007} A. Hucht, Phys. Rev. Lett. \textbf{99}, 185301 (2007).

\bibitem{Vasilyev2007} O. Vasilyev, A. Gambassi, A. Maciołek, and S. Dietrich, Europhys. Lett. \textbf{80}, 60009 (2007).

\bibitem{Maciolek2007} A. Maciolek, A. Gambassi, and S. Dietrich, Phys. Rev. E \textbf{76}, 031124 (2007).

\bibitem{Hasenbusch2009} M. Hasenbusch, J. Stat. Mech. P07031 (2009).

\bibitem{Hasenbusch2010} M. Hasenbusch, Phys. Rev. B \textbf{81}, 165412 (2010).

\bibitem{Garcia1999} R. Garcia and M. H. W. Chan, Phys. Rev. Lett. \textbf{83}, 1187 (1999).


\bibitem{Ganshin2006} A. Ganshin, S. Scheidemantel, R. Garcia, and M. H. W. Chan, Phys. Rev. Lett. \textbf{97}, 075301 (2006).

\bibitem{Aydiner2013} E. Aydiner, (under review, 2014).


\bibitem{Molina1996}  M. I. Molina, Am. J. Phys. \textbf{64}, 503 (1996).


\bibitem{Pathira1998}  R. K. Pathira, Am. J. Phys. \textbf{66}, 1080 (1998).

\bibitem{Dai2003}  W. S. Dai  and M. Xie, Phys. Lett. A \textbf{311}, 340 (2003).

\bibitem{Dai2004}  W. S. Dai  and M. Xie, Phys. Rev. E \textbf{70}, 016103 (2004).

\bibitem{Sisman2004}  A. Sisman and I. Muller, Phys. Lett. A \textbf{320}, 360 (2004).


\bibitem{Pang2006}  H. Pang, W. S. Dai and M. Xei, J. Phys. A: Math. Gen. \textbf{39}, 2563 (2006).

\bibitem{Firat2013}  C. Firat and A. Sisman, Phys. Scr. \textbf{87}, 045008 (2013).

\bibitem{Nie2008a} W. J. Nie, J. Z. He and X. J. He, J. Appl. Phys. \textbf{103}, 114909 (2008).

\bibitem{Nie2008b} W. J. Nie and J. Z. He, Phys. Lett. A \textbf{372}, 1168 (2008).

\bibitem{Nie2009a} W. J. Nie, J. Z. He and J. Du, Physica A \textbf{388}, 318 (2009).

\bibitem{Nie2009b} W. J. Nie, J. Z. He and J. Du, J. Appl. Phys. \textbf{105}, 054903 (2009).

\bibitem{Nie2010} W. J. Nie, Q. Liao, C. Zhang and X. J. He, Energy \textbf{35}, 4658 (2010).

\bibitem{Greiner2002} M. Greiner, O. Mandel, T. Esslinger, T. W. H\"{a}nsch and I. Bloch, Nature
\textbf{415}, 39 (2002)

\bibitem{Bargi2006}  S. Bargi, G. M. Kavoulakis, S. M. Reimann, Phys. Rev. A
\textbf{73}, 033613 (2006).

\bibitem{Cozzini2006}  M. Cozzini, B. Jackson, S. Stringari, Phys. Rev. A
\textbf{73}, 013603 (2006).

\bibitem{Fetter2001} A. L. Fetter, Phys. Rev. A \textbf{64}, 063608 (2001).

\bibitem{Ghosh2004} T. K. Ghosh, Phys. Rev. A \textbf{69}, 043606 (2004).

\bibitem{Kim2005} J. K. Kim, A. L. Fetter, Phys. Rev. A \textbf{72}, 023619 (2005).

\bibitem{Danaila2005} I. Danaila, Phys. Rev. A \textbf{72}, 013605 (2005.)

\bibitem{Fetter2005}  A. L. Fetter, B. Jackson, S. Stringari, Phys. Rev. A \textbf{71}, 013605 (2005).

\bibitem{Kling2007} S. Kling, A. Pelster, Phys. Rev. A \textbf{76}, 023609 (2007).


\bibitem{Blakie2007} P. B. Blakie, A. Bezett, P. F. Buonsante, Phys. Rev. A \textbf{75}, 063609 (2007).

\bibitem{Dupuis2009} N. Dupuis, K. Sengupta, Physica B \textbf{404}, 517 (2009).

\bibitem{Gerbier2008} F. Gerbier et al., Phys. Rev. Lett. \textbf{101}, 155303 (2008).

\bibitem{Hassan2010} A. S. Hassan, Phys. Lett. A  \textbf{374}, 2106 (2010).

\bibitem{Ketterle1996} W. Ketterle, N. J. van Druten, Phys. Rev. A  \textbf{54}, 656 (1996).


\bibitem{Hunter2000} J. Hunter and B. Nachtergaele, Applied Analaysis (World Scientific, 2000).



\end{thebibliography}

\end{document}